\documentstyle[12pt]{article}
\def\baselinestretch{1.4}
\def\lab#1      {\hbox{\small #1} }
\newcommand{\ben}{\begin{eqnarray*}}
\newcommand{\een}{\end{eqnarray*}}
\newcommand{\benn}{\begin{eqnarray}}
\newcommand{\eenn}{\end{eqnarray}}

\def\be           {\begin{equation}}
\def\ee           {\end{equation}}

\def\Tr           {\mathop{\hbox{Tr}}}
\def\mb#1         {\mbox{\boldmath $#1$}}
\def\diffn#1	  {\Delta^{-}_{#1}}

\def\la           {\left\langle }
\def\ra           {\right\rangle }
\def\lhs          {{\sc lhs} }

\def\e 		  {\right|_{\varepsilon = 0}}
\def\none	  {\multicolumn{2}{c}{---}}

\def\notesize	  {\tiny }

\def\static	  {\hbox{\notesize (static)}}
\def\dyn	  {\hbox{\notesize (dyn)}}
\def\gauge	  {\hbox{\notesize (gauge)}}
\def\total	  {\hbox{\notesize (total)}}
\def\fp		  {\hbox{\notesize (FP)}}

\begin{document}

\begin{titlepage}

\begin{tabbing}
\` {\sl hep-lat/9807001} \\
    \\
\` LSUHE No. 269-1998 \\
\` June, 1998 \\
\end{tabbing}
 
\vspace*{0.1in}
 
\begin{center}
{\large\bf Ehrenfest theorems and charge antiscreening
in Abelian projected gauge theories\\}
\vspace*{.5in}
Giuseppe Di Cecio$^1$, 
Alistair Hart$^2$
and Richard W. Haymaker$^3$\\
\vspace*{.2in}
{\small\sl Department of Physics and Astronomy, Louisiana State
University,\\ Baton Rouge, Louisiana 70803-4001, U.S.A.
\\
\vspace{3ex}
$^1$Present address:
Specialised Derivatives Section, Midland Bank plc., \\
Thames Exchange, 10 Queen Street Place, London EC4R 1BQ, U.K. 
\\
\vspace{2ex}
e--mail addresses:      $^2$hart@rouge.phys.lsu.edu, 
                  $^3$haymaker@rouge.phys.lsu.edu}
\end{center}

\vspace{0.2in}
\begin{center}
\begin{minipage}{5in}
\begin{center}
{\bf Abstract.}
\end{center}

We derive exact relations for SU(2) lattice gauge theory in 3+1 
dimensions. In terms of Abelian projection, these are the
expectation values of Maxwell equations that define a new field
strength operator and conserved, dynamic electric currents formed 
from the charged matter and ghost fields. 
The effect of gauge fixing is calculated, 
and in the maximally Abelian gauge we find antiscreening of
U(1) Wilson loop source charges. We discuss the importance of
these quantities in the dual superconducting vacuum mechanism of
confinement.

\end{minipage}
\end{center}

\vfill
\noindent
PACS indices: 
11.15.Ha, 
11.30.Ly. 

\end{titlepage}

\setcounter{page}{1}
\newpage
\pagestyle{plain}

\section{Introduction}
\label{sect_intro}

Lattice studies based on Abelian projection have had 
considerable success identifying
the dynamical variables relevant to the physics of
quark confinement. There is no definitive way as yet
of choosing the optimum variables, but in the 
maximally Abelian gauge 
\cite{thooft81,kronfeld87a}
the U(1) fields remaining after Abelian projection
produce a heavy quark potential that continues to rise
linearly 
\cite{suzuki90}.
Further the string tension is almost, but not exactly,
equal to the full SU(2) quantity; 92\% in a recent study
at $\beta = 2.5115$
\cite{bali96}.

This suggests that we may be close to identifying an
underlying principle governing confinement.
All elements of a dual superconducting vacuum appear to
be present 
\cite{mandelstam76,thooft81};
in the maximally Abelian gauge  
magnetic monopoles reproduce nearly all of the U(1) string
tension
\cite{stack94,bali96}.
The spontaneous breaking to the U(1) gauge symmetry 
is signalled by the
non-zero vacuum expectation value of monopole operator
\cite{chernodub96,digiacomo98}.
The profile of the electric field and the 
persistent magnetic monopole currents 
in the vortex between quark and antiquark  
are well described by an
effective theory, the Ginzburg--Landau, or equivalently a Higgs
theory giving a London penetration depth and Ginzburg--Landau 
coherence length
\cite{singh93,bali98}.

Central to finding the effective theory is the definition of
the field strength operator in the Abelian projected theory,
entering not only in the vortex profiles but also in the
formula for the monopole operator. All definitions should be
equivalent in the continuum limit, 
but use of the appropriate lattice
expression should lead to a minimisation of discretisation
errors.

In this paper we exploit lattice symmetries to derive such an
operator that satisfies Ehrenfest relations; Maxwell's 
equations for ensemble averages irrespective of
lattice artefacts. In section~\ref{sec_u1} we introduce and
review this method in pure U(1) theories
\cite{zach95},
discussing the Abelian projected SU(2) theory, with and without
gauge fixing to the maximally Abelian gauge, in 
section~\ref{sec_nonab}.

The charged coset fields are normally discarded in Abelian
projection, as are the ghost fields arising from the 
gauge fixing procedure. Since the remainder of the SU(2)
infrared physics must arise from these, an understanding of
their r{\^o}le is central to completing the picture of full
SU(2) confinement. In section~\ref{sec_nonab} we begin
to address this issue, showing that these fields form a
charged U(1) current, and in section~\ref{sec_numerics}
demonstrate that in the maximally Abelian gauge the
supposedly unit charged Abelian Wilson loop has an
upward renormalisation of charge due to this current 
of $O(15\%)$. A localised cloud of like polarity charge is
induced in the vacuum in the vicinity of a source, producing
an effect reminiscent of the antiscreening of charge in 
{\sc QCD}. In other gauges studied, the analogous current
is weaker, and acts to {\em screen} the source.

We show that this current can be quantitatively written as
a sum of terms from the coset and ghost fields. The contribution
of the ghost fields in the maximally Abelian gauge in
this context is found to be small. The effect of the
the Gribov ambiguity on these currents is argued to
be  slight.

Finally in section~\ref{sec_summ} we discuss these results. 
Some preliminary results have already appeared
\cite{dicecio97}.
The reader's attention is drawn to related work in this subject
\cite{suganuma97,skala97}.

\section{Abelian theories}
\label{sec_u1}

The Wilson action
\be
S_W = \sum_{n,\mu < \nu} (1 - \cos \theta_{\mu \nu}(n))
\ee
comprises link angles 
$
\{ \theta_\mu(n) \in [-\pi,\pi) \}  
$
summed to form plaquette angles 
$
\theta_{\mu \nu}(n) =  
\theta_\mu(n) + \theta_\nu(n+\hat{\mu}) - 
\theta_\mu(n+\hat{\nu}) - \theta_\nu(n).
$ 
External electric sources may be represented
by a Wilson loop, and we consider here for
simplicity a specific plaquette on the lattice
$ P_{\kappa \lambda} (n^\prime) =
\exp i\theta_{\kappa \lambda} (n^\prime)$ 
with real and imaginary parts 
$R_{\kappa \lambda}$, $I_{\kappa \lambda}$
respectively. The partition function
\be 
Z_{\cal SRC} = \int [d \theta_\mu] 
P_{\kappa \lambda} (n^\prime) 
e^{-\beta S_W}.
\ee
is invariant under the introduction of an
arbitrary constant into any link angle.  We call this a shift
invariance and we focus on the consequence of this on a
particular link: $\theta_\mu(n) \rightarrow \theta_\mu(n) +
\varepsilon$. The shift
corresponds to either a left or right multiplication of that link by a
group element:
$
e^{i \theta_\mu(n)} \rightarrow 
e^{i \theta_\mu(n)} e^{i \varepsilon} =
e^{i \varepsilon} e^{i \theta_\mu(n)}.
$
The Haar measure  is invariant under this transformation:
$
d (\theta_\mu(n) + \varepsilon) = d \theta_\mu(n)
$.

The first order shift in the partition function gives an identity:
\benn
\left.
\frac{1}{Z_{\cal SRC}}
\frac{\partial Z_{\cal SRC}}{\partial \varepsilon}
\e
&=& 
\frac{1}{Z_{\cal SRC}}
\int [d \theta_\mu] 
\left[
\left.
 \frac{\partial P_{\kappa \lambda}(n^\prime)}{\partial \varepsilon} \e
-
P_{\kappa \lambda}(n^\prime) .
\beta
\left. \frac{\partial S_W}{\partial \varepsilon} \e
\right]
e^{-\beta S_W} 
\nonumber \\ 
\nonumber \\
&=& 
\frac{1}{\int [d \theta_\mu] 
R_{\kappa \lambda} .
e^{-\beta S_W}   }
\int [d \theta_\mu] 
\left[
i R_{\kappa \lambda} . \delta_W
- 
i I_{\kappa \lambda} . 
\beta
\left.
\frac{\partial S_W}{\partial \varepsilon} \e
\right]
e^{-\beta S_W} 
\nonumber \\ 
\nonumber \\
&=& 
i \delta_W
- \frac{i \beta}{\la R_{\kappa \lambda} \ra}
\la
I_{\kappa \lambda} .
\left. \frac{\partial S_W}{\partial \varepsilon} \e
\ra = 0
\eenn
where $\delta_W \neq 0 $ only when the shifted link coincides with a
link in the source,
\be
\delta_W =
  \left[
\delta_{\mu\kappa}
(\delta_{n,n^\prime} - \delta_{n,n^\prime+\hat{\lambda}}) -
\delta_{\mu\lambda} 
(\delta_{n,n^\prime} - \delta_{n,n^\prime+\hat{\kappa}})
\right].
\label{eqn_deltaW}
\ee
The real or imaginary part of $P_{\kappa \lambda}$ is dropped if it
contributes a term odd in the link angle and hence has a zero expectation 
value.

Multiplying eqn.~(\ref{eqn_deltaW}) by the electric charge $e$, where $ \beta
= 1/e^2$, we use the backwards lattice difference operator to define
the lattice field strength tensor, $f_{\mu \nu}$
\be 
\frac{1}{e}
\left. \frac{\partial S_W}
{\partial \varepsilon} 
\right|_{\varepsilon = 0}
=  \frac{1}{e} \diffn{\nu} I_{\mu \nu} (n)
\equiv    \diffn{\nu} f_{\mu \nu} (n).
\ee
  The identity becomes:
\be
\frac{
\la 
I_{\kappa \lambda} . 
\diffn{\nu} f_{\mu \nu} (n)
\ra}
{
\la 
R_{\kappa \lambda}
\ra
}
=
j^{\static}_\mu(n)
\ee
where the static current density is
\be
j^{\static}_\mu(n)
 =  e \delta_W
\ee

We have arrived at what appear to be a discretised version of the
continuum Maxwell's equations for the U(1) fields, but satisfied by
the expectation values rather than merely in the classical limit of
extremising the action.  In quantum mechanics, Ehrenfest's theorem
relates the time derivative of the position operator to the potential
in a way reminiscent of Newton's classical equations of motion,
$
m \frac{d^2}{dt^2} \la \hat{\mb{x} } \ra = 
- \la \mb{\nabla} V(\hat{\mb{x} }) \ra. 
$
By analogy, we label the lattice expressions `Ehrenfest identities'%
\footnote{%
A term also used in Zach et. al.
\cite{zach95}.
}.

The theorem defines the lattice field strength operator
whose form is dictated by the derivative of the 
U(1) lattice action. As all actions (in the same universality class) 
are equivalent in the weak coupling limit, so too are the 
corresponding definitions of the field strength operator. Much work,
however, is performed at finite lattice spacing, and it is 
advantageous to avoid extraneous $O(a^2)$ effects by using the 
`correct' operator. Using this operator we may measure the charge
density, which the theorem shows is exactly that of the introduced 
source. No further charge is induced.

\section{Non--Abelian gauge theories}
\label{sec_nonab}

The correct Abelian field strength operator could be calculated as
above if the effective Abelian action for the U(1) fields after
Abelian projection were known. It is not, and we approach the
problem from the full SU(2) action. The SU(2) theory has symmetries
analogous to those of the pure Abelian theory, each of which gives
rise to Ehrenfest identities. 
Since these, in the continuum limit, resemble
the Euler--Lagrange equations for the corresponding continuum
action, we begin by briefly considering these.

The continuum Lagrangian is
$
{\cal L} =
\frac{1}{4} G^a_{\mu \nu}(x) G^a_{\mu \nu}(x)
$,
the isospin index $a \in \{ 1,2,3 \}$. Under Abelian
projection  the third component of the gauge field becomes
 the Abelian gauge potential. We can 
rewrite ${\cal L}$ to emphasis this:
\be
{\cal L} = \frac{1}{4} \left(
F_{\mu \nu} F_{\mu \nu} + W^{*}_{\mu \nu} W_{\mu \nu} 
+ \frac{i}{2} F_{\mu \nu} \left( 
W_\mu W^{*}_\nu -  W^{*}_\mu W_\nu \right)
- \frac{1}{4} \left(
W_\mu W^{*}_\nu -  W^{*}_\mu W_\nu \right)^2
\right)
\ee
where $F_{\mu \nu}(x) = \partial_\mu
A^3_{\nu}(x) - \partial_\nu A^3_{\mu}(x)$ is the Abelian field 
strength, the remaining components forming a complex matter 
field $W_\mu(x) = A^1_{\mu}(x) + i A^2_{\mu}(x)$. This field is 
electrically charged with respect to the photon; 
$D_\mu W_\nu(x) = (\partial_\mu - iA^3_\mu(x)) W_\nu(x)$,
giving $W_{\mu \nu}(x) = D_\mu W_\nu(x) - D_\nu W_\mu(x)$.

Consider the extremisation with respect to
$A^3_\mu(x)$. This gives what appear to be Maxwell's equations
with a dynamical, real--valued, conserved electric current formed from
the coset (matter) fields and their coupling to the photon.
\be
\partial_\nu F_{\mu \nu}(x) = J^{\dyn}_\mu(x) = 
\frac{-i}{4} \left( 2 \left(
W_{\mu \kappa} W^{*}_\kappa - W^{*}_{\mu \kappa} W_\kappa
\right)
- \partial_\kappa \left(
W_\kappa W^{*}_\mu - W^{*}_\kappa W_\mu 
\right)
\right).
\label{eqn_continuum_EL}
\ee
The first two terms in the current are precisely what would be expected 
for a charged vector field, i.e. where $D_\mu W_\mu(x) = 0$. This
is not in general true. It is interesting to note, however, that the 
imposition of this constraint amounts precisely to fixing the theory
to the maximally Abelian gauge.

The derivation here relied upon the assumption that we might vary the
photon field independently of the charged coset fields. While this is
true in the full theory, a gauge fixing constraint couples
the variations in the fields. Such an extremisation problem is
usually tackled using Lagrange multipliers. The correct
lattice operators cannot be predicted by na\"{\i}ve discretisation of
continuum results, so we do not pursue this approach here but
move on to the lattice Ehrenfest identities.

\subsection{Abelian projection on the lattice}

After gauge fixing, the SU(2) link matrices may be decomposed in a 
`left coset' form:
\be
 U_\mu (n) = \left (
\begin{array}{cc}
\cos (\phi_\mu (n))  &
\sin (\phi_\mu (n)) e^{i\gamma_\mu (n)} \\& \\
-\sin (\phi_\mu (n)) e^{ -i\gamma_\mu (n)} &
\cos (\phi_\mu (n))  \\
\end{array}
\right )
\left (
\begin{array}{cc}
e^{ i\theta_\mu (n)} & 0 \\
\\
0 & e^{ -i\theta_\mu (n)} \\
\end{array}
\right ), 
\ee
Under a U(1) gauge transformation, 
$ \left\{ g(n) = \exp \left[ i \alpha(n) \sigma_3 \right] \right\} $,
\be
\theta_{\mu}(n) \rightarrow \theta_{\mu}(n) + \alpha(n) -
\alpha(n + \hat{\mu}) 
\mbox{\hspace{2em}}
\gamma_{\mu}(n) \rightarrow \gamma_{\mu}(n) + 2 \alpha(n)
\ee
In other words, the left coset field derived from the
link $U_\mu(n)$ is a doubly charged matter field living on the
site $n$ and is invariant under U(1) 
gauge transformations at neighbouring sites.  

The $c_\mu \equiv \cos (\phi_\mu)$ 
are real--valued fields which near the continuum $\sim 1
+ O(a^2)$ where $a$ is the lattice spacing. The off--diagonal 
$w_\mu \equiv \sin (\phi_\mu) e^{i\gamma_\mu}$
become the charged coset fields $g a W_\mu(x)$, and $\theta_\mu$ the
photon field $g a A^3_\mu(x)$. [The SU(2) coupling $\beta =
\frac{4}{g^2}$ in 3+1 dimensions.]

The SU(2) shift symmetries are the left and right multiplications of a
link by an arbitrary constant SU(2) matrix, under which the Haar
measure is invariant. Since it is the $a=3$
component that becomes the photon, we consider here only shift matrices of
the form 
$ \bar{U} = \exp [i \varepsilon \sigma_3] $.
\be
\begin{array}{rclcll}
{\hbox{right}} & : & U_\mu(n) \rightarrow U_\mu(n) \bar{U} 
& \equiv &
\left\{
w_\mu(n) \rightarrow w_\mu(n) \right., &\left.
\theta_\mu(n) \rightarrow \theta_\mu(n) + \varepsilon
\right\}
\\
{\hbox{left}}  & : & U_\mu(n) \rightarrow \bar{U} U_\mu(n) 
& \equiv &
\left\{
w_\mu(n) \rightarrow w_\mu(n) e^{i 2 \varepsilon} \right., &
\left.
\theta_\mu(n) \rightarrow \theta_\mu(n)+ \varepsilon
\right\} .
\label{eqn_su2_shifts}
\end{array}
\ee

\subsection{The identities -- no gauge fixing}
\label{sec_su2_no}

We first derive the Ehrenfest relations in the simpler,
but artificial, context of Abelian projection without 
gauge fixing, `no gauge.'
The SU(2) link matrices combine to form plaquettes 
$ 
U_{\mu \nu}(n) = U_\mu(n) U_\nu(n+\hat{\mu})
U^\dagger_\mu(n+\hat{\nu}) U^\dagger_\nu(n)
$
and the Wilson action
\be
S_W = \sum_{n,\mu < \nu} 
\left( 1 - \frac{1}{2} \Tr U_{\mu \nu}(n) \right).
\ee
Writing each link as the sum of a diagonal and an off--diagonal
matrix, $U = D + O$, we define the product of the diagonal terms
around a plaquette
\be
Q_{\kappa \lambda}(n^\prime) = 
D_\kappa(n^\prime) D_\lambda(n^\prime+\hat{\kappa})
D^\dagger_\kappa(n^\prime+\hat{\lambda}) D^\dagger_\lambda(n^\prime).
\ee
and choose as a source term a specific plaquette $P_{\kappa
\lambda}(n^\prime)$ with real and imaginary parts:
\benn
R_{\kappa \lambda} \equiv
\frac{1}{2} \Tr \{Q_{\kappa \lambda}(n^\prime) \} & = &
\left( 
c_\kappa(n^\prime) . c_\lambda(n^\prime+\hat{\kappa}) . 
c_\kappa(n^\prime+\hat{\lambda}) . c_\lambda(n^\prime)
\right)
\cos \theta_{\kappa \lambda}(n^\prime), 
\nonumber \\
I_{\kappa \lambda} \equiv
\frac{1}{2} \Tr \{ i \sigma_3 Q_{\kappa \lambda}(n^\prime) \} & = &
\left( 
c_\kappa(n^\prime) . c_\lambda(n^\prime+\hat{\kappa}) . 
c_\kappa(n^\prime+\hat{\lambda}) . c_\lambda(n^\prime)
\right)
\sin \theta_{\kappa \lambda}(n^\prime).
\eenn
It is for consistency that we include the coset fields.
The fields $c_\kappa(n^\prime) \rightarrow 1$ in the continuum limit
and in the maximally Abelian gauge can be considered to be almost
constant
\cite{chernodub95,poulis96}, 
since the fluctuations are very small. For this reason, we anticipate
only minor changes in the measured averages if the fields 
$c_\mu(n)$ were excluded from the source as is the case in the
traditional Abelian source loop.

The partition function is
\be
{\cal Z}_{{\cal SRC}} = \int [d U_\mu] 
P_{\kappa \lambda}(n^\prime)
e^{-\beta S_W}
\ee
If all fields bar the $\theta_\mu$ are discarded, the SU(2)
action reduces to the U(1) Wilson action, so
we define an effective electric charge, $e$, by 
$\beta = \frac{1}{e^2}$. 
The shift invariance gives an analogous result to the
U(1) theory
\benn
j^{\static}_\mu(n)
- \frac{e \beta}{\la R_{\kappa \lambda} \ra}
\la
I_{\kappa \lambda}
\left. \frac{\partial S_W}{\partial 
\varepsilon} \e
\ra = 0
\eenn
and the static current density is also based on the localised form of
eqn.~(\ref{eqn_deltaW}):
$
j^{\static}_\mu(n)
 =  e \delta_W.
$
Although we have introduced an Abelian projected source, the links, $U_\mu$,
in the derivative of the action are SU(2) matrices. Were we to replace them
also with their diagonal components only, 
$D_\mu = \mbox{diag } (c_\mu e^{i \theta_\mu}, c_\mu e^{-i \theta_\mu})$,
we should get a result analogous to the U(1) case:
\be
\left.
\frac{1}{e} 
 \frac{\partial S_W}
{\partial \varepsilon} 
\right|_{\varepsilon = 0; U \rightarrow D}
= \frac{1}{e}  \diffn{\nu} I_{\mu \nu}(n)
= \diffn{\nu} f_{\mu \nu}(n)
\ee
The remaining terms involving both $D$ and $O$ 
constitute a U(1) gauge invariant dynamical current.
\be
\left.
\frac{1}{e} 
 \frac{\partial S_W}
{\partial \varepsilon} 
\right|_{\varepsilon = 0}
= \diffn{\nu} f_{\mu \nu}(n) -  j^{\dyn}_\mu(n)
\label{eqn_nonab_dSw}
\ee
The terms in $j^{\dyn}_\mu$ that survive into the continuum
limit yield the continuum Euler--Lagrange result (eqn.~(\ref
{eqn_continuum_EL})).
The Ehrenfest identity is thus:
\be
\frac{
\la 
I_{\kappa \lambda}  . 
\diffn{\nu} f_{\mu \nu} (n)
\ra
}
{
\la 
R_{\kappa \lambda}
\ra
}
=
j^{\static}_\mu(n)
+
\frac{
\la 
 I_{\kappa \lambda} .  
j^{\dyn}_\mu(n)
\ra
}
{
\la 
 R_{\kappa \lambda}
\ra
}
\label{eqn_ehren_nonab}
\ee

\subsection{Gauge fixing}

We restrict the discussion here to the 
maximally Abelian gauge, defined as the local maximisation
by gauge transformations 
$\{ g(n) = \exp [i \alpha_a(n) \sigma_a ] \} $
of the Morse functional
\be
R[U] = - \sum_{n,\mu} \Tr \left\{
i \sigma_3 . U_\mu(n) . i \sigma_3 . U^\dagger_\mu(n)
\right\}.
\ee
Denoting the gauge transformed link variable as
$
U^g_\mu (n) \equiv g(n) U_\mu(n) g^\dagger (n+\hat{\mu}) 
$,
at a local maximum, assumed to occur at $g = 1$, 
the first derivative with respect to gauge
transformations of $R[U^g]$ will be zero: 
\be
F^r[U;p] \equiv  
\left.
\frac{\partial R[U^g]}{\partial \alpha_r(p)}
\right|_{\alpha = 0}
=
-2 \epsilon^{3rc} \Tr \left\{ i \sigma_c . X(n) \right\} = 0
\ee
where $X(p) =
\sum_{\mu > 0} \left(
U_\mu(p).i \sigma_3.U^\dagger_\mu(p) +
U^\dagger_\mu(p-\hat{\mu}).i \sigma_3.U_\mu(p-\hat{\mu})
\right).
$
The residual U(1) symmetry is seen in $F^3$ being trivially zero.
The gauge fixed partition function is thus
\be
{\cal Z}^{gf}_{\cal SRC} = \int [d U_\mu] P_{\kappa \lambda}[U;n^\prime] .
e^{-\beta S_W[U]} . \Delta_{FP}[U] \left(
\prod_{p,r} \delta \left( F^r[U;p] \right) \right)
\label{eqn_nonabZ}
\ee
and $\Delta_{FP}$ is the Faddeev--Popov gauge fixing operator, a 
Jacobian to reweight the integration measure after the introduction
of the constraint into the partition function:
\be
\Delta_{FP}[U] = 
\left(
\int [dg] \prod_{p,r} \delta \left( F^r[U^g;p] 
\right) \right)^{-1} = 
\left| \det M^{rs}_{pq} \right|
\mbox{\hspace{2em} and } M^{rs}_{pq} \equiv \left.
\frac{\partial F^r[U^g;p]}{\partial \alpha_s(q)}
\right|_{\alpha = 0}
\ee
is the second term in the Taylor expansion of $R$
\cite{zwanziger97}.
The integral expression is gauge invariant by the invariance 
of the Haar measure. The
determinant is gauge invariant by fiat; we must evaluate it `on
the constraint' (i.e. at $F=0$). If $F \not= 0$ we must first move
the configuration along the gauge orbit by gauge transformations
until it does.

Applying a shift to a single link and differentiating with respect to the shift 
parameter, $\varepsilon$, now raises problems, as the shift drives the 
previously gauge fixed configuration off the constraint. 
This manifests as derivatives of the 
constraint $\delta$--functions, which we must avoid. One approach would
be to shift not one link, but each of a Polyakov line of links extending
around the lattice. By careful choice of the relative sizes of these shifts
we may conspire to remove the offending $\delta$--function
terms using the constraint equations. This is at the expense of the shift as 
a local probe of charge density. Instead we accompany the shift of one specific 
link by
an SU(2) gauge transformation over the whole lattice that is also linear in
the shift parameter. The combined effect of the shift and `corrective' gauge 
transformation is to move on a path in configuration space parameterised by
$\varepsilon$ that remains 
on the trajectory satisfying the maximally Abelian gauge. There is 
potentially also a non--locality here, but we shall see that it
is very limited in its extent.

The corrective gauge transformation
$\left\{ g(q) = 1 + i \varepsilon \eta_s(q) \sigma_s \right\}$ is calculated
using the inverse of the Faddeev--Popov matrix, and so $\eta_3 = 0$:
\be
\eta_s(q) = - \sum_{p,r} (M^{-1})^{sr}_{qp}
\left. \frac{\partial F^r[U^\varepsilon;p]}{\partial \varepsilon} \e
\ee
We now apply both the shift and corrective gauge transformation to the 
partition function in eqn.~(\ref{eqn_nonabZ}), and differentiate with respect to
$\varepsilon$. The action is gauge invariant, and thus only the shift has an
effect and the field strength operator and current $j^{\dyn}_\mu$ are 
as before (eqn.~(\ref{eqn_nonab_dSw})).

The derivative of the source plaquette now makes two 
contributions; the
shift gives the same static current as before.
The second comes from the corrective gauge transformation, and
forms a conserved current $j^{\gauge}_\mu$.

Finally, there are contributions that arise from perturbing the links 
making up the Faddeev--Popov operator. Under this, the matrix 
$M \rightarrow M + \varepsilon N$. 
As we have remarked, the Faddeev--
Popov operator is specifically 
the determinant of $M$ {\em evaluated on the constraint.} By introducing
$\eta$ we have stayed on the constraint as we shifted the link, and 
$N = A + B$ has contributions from the shift, $A$, and from the corrective 
gauge
transformation, $B$. The derivative forms a further conserved current:
\be
j^{\fp}_\mu = \frac{e}{\Delta_{FP}[U]} \left.
\frac{\partial \Delta_{FP}[U^\varepsilon]}{\partial \varepsilon}
\e
= e
\left.
\frac{\partial \det(1 + \varepsilon M^{-1} N)}
{\partial \varepsilon}
\e
=
e \Tr \left\{ M^{-1} N \right\}
\ee
Defining $j^{\total}_\mu = j^{\dyn}_\mu + j^{\gauge}_\mu + j^{\fp}_\mu$, 
we have as the final result a set of identities
\be
\frac{
\la 
I_{\kappa \lambda}  . 
\diffn{\nu} f_{\mu \nu} (n)
\ra_{gf}
}
{
\la 
R_{\kappa \lambda}
\ra_{gf}
}
=
j^{\static}_\mu(n)
+
\frac{
\la 
 I_{\kappa \lambda} .  
j^{\total}_\mu(n)
\ra_{gf}
}
{
\la 
 R_{\kappa \lambda}
\ra_{gf}
}
\label{eqn_ehren_nonabgf}
\ee
where the gauge fixed expectation value is defined as
\be
\la {\cal O} \ra_{gf} \equiv \int [d U_\mu] 
.{\cal O}.
e^{-\beta S_W[U]} . \Delta_{FP}[U] \left(
\prod_{p,r} \delta \left( F^r[U;p] \right) \right).
\ee
We conclude this section with two remarks.
In simulation, a lattice average using SU(2) configurations
calculated using standard Monte Carlo methods and then
gauge fixed will include the Faddeev--Popov
operator in the importance sampling of the measure 
and it need not be explicitly calculated.

Secondly, when deriving the Faddeev--Popov matrix, there
is some ambiguity as to whether one should start from the
constraint
\be
M^{rs}_{pq} = 
\left.
\frac{\partial F^r[U^g;p]}{\partial \alpha_s(q)}
\right|_{\alpha = 0} = 
\left.
\frac{\partial}{\partial \alpha_s(q)}
\left(
\left.
\frac{\partial R[U^g]}{\partial \alpha_r(p)}
\right|_{\alpha = 0}
\right)
\right|_{\alpha = 0}
\ee
or from the original functional
\be
M^{rs}_{pq} = 
\left.
\frac{\partial^2 R[U^g]}{\partial \alpha_s(q) \partial \alpha_r(p)}
\right|_{\alpha = 0}.
\ee
Since we believe that $M$ should be a symmetric matrix in general,
the latter seems the most natural. In practise, we find that the
differences between the two approaches are all multiples of the
constraint $F$ and hence when the gauge condition is satisfied
the two definitions coincide. 
Similar ambiguities arise when the response 
of the Faddeev--Popov matrix to the corrective gauge 
transformation is considered,
since for this we require the third term, $L$, in the Taylor expansion 
of $R$ 
\be
B^{rs}_{pq} = \sum_{u,t} L^{rst}_{pqu} \eta_t(u)
\ee
and it is unclear whether we should begin from $R$, $F$ or $M$ in
its derivation.
Again we find that the most symmetric tensor is derived from $R$,
but that all expressions are the same on the constraint. This we
believe to be a general property.

The ambiguities are more serious in the case of the contribution
of the shift to the Faddeev--Popov matrix. The matrix $A$ is 
derived from $R$ by two differentiations with respect to
gauge transformations, and one with respect to the shift. The
order of these, and at what point variables are set to zero
does appear to matter in this case, even when $F=0$. 
The correct order of derivatives is as follows:
\be
A^{rs}_{pq} =  
\left.
\frac{\partial}{\partial \alpha_s(q)}
\left(
\left.
\frac{\partial}{\partial \varepsilon}
\left(
\left.
\frac{\partial R[U^g]}{\partial \alpha_r(p)}
\right|_{\alpha = 0}
\right)
\e
\right)
\right|_{\alpha = 0}.
\ee
The inner two nested derivatives give the first order shift correction
to the constraint, i.e., the second term in $F + \epsilon F_{\epsilon} = 0$. 
The Faddeev-Popov operator is the matrix of derivatives of this about the
constraint and is formed by the outermost nested derivative.  
The corrected Faddeev-Popov operator must be calculated about the
shifted constraint. 
If, for example,
 we reverse
the order of the outer two nested derivatives we would be calculating the
shift of the lowest order Faddeev-Popov operator.  This would give an incorrect
contribution to the current.  Numerical studies confirms these conclusions. 
\section{Numerical investigation}
\label{sec_numerics}

The terms in the Ehrenfest identities may be measured using Monte
Carlo simulation. We begin with a careful test of the expressions,
and for this we rewrite 
eqns.~(\ref{eqn_ehren_nonab},\ref{eqn_ehren_nonabgf})
to minimise the combination of statistical errors:
\be
\begin{array}{l}
\beta
\la 
I_{\kappa \lambda}  . 
\left.
\frac{\partial S_W}{\partial \varepsilon}
\e
\ra
-
k^{\static}_\mu(n) . \la R_{\kappa \lambda} \ra
= 0
\\
\beta
\la 
I_{\kappa \lambda}  . 
\left.
\frac{\partial S_W}{\partial \varepsilon}
\e
\ra_{gf}
-
k^{\static}_\mu(n) . \la R_{\kappa \lambda} \ra_{gf}
-
\la I_{\kappa \lambda} . k^{\gauge}_\mu(n) \ra_{gf}
-
\la I_{\kappa \lambda} . k^{\fp}_\mu(n) \ra_{gf}
= 0
\label{eqn_ehren_test}
\end{array}
\ee
where the currents $k$ differ from $j$ by a factor of the charge.
We further split $k^{\fp}_\mu$ into the (normally summed)
contributions from the shift and corrective gauge
transformation. To measure the gauge fixed currents separately
requires the inversion of the large Faddeev--Popov matrix. This
limits us to the unphysical lattice size of $4^4$ for this test,
where although the identities must still hold exactly, there may
be significant finite volume effects on the individual currents.
For comparison we also test the `no gauge' expression on the same
lattice. 

In Table~\ref{tab_oldsource} we show results where the (left) 
shifted link is part of the source plaquette 
(i.e. $|k^{\static}_\mu| = 1$).
The identities are verified within very small statistical errors.
The numbers are for right shifts are identical within these errors.
Off the source, the identities are equally well satisfied.

Table~\ref{tab_onandoff} shows the normalised currents arising from
shifting a timelike link included in the source, and then one 
spacelike lattice spacing away from the source in the same plane. In
the maximally Abelian gauge we see that on the source the charge of
the Abelian projected Wilson loop has been renormalised from unity to
a value $O(15\%)$ higher. Further, in the vicinity of the source we 
find  a cloud of like charges has been induced in the vacuum. This is
reminiscent of the charge antiscreening (or asymptotic freedom) of the
full gauge theory. The charge cloud is localised, falling to near zero
by two lattice spacings from the source. We note that in the case of
`no gauge' (and also in the Polyakov and diagonal plaquette gauges
e.g. `$F_{\mu \nu}$') there is also a renormalisation, but that it
reduces the charge of the Abelian Wilson loop. The induced charge
cloud is weaker and acts to screen the source, at odds with
the behaviour of the full non--Abelian theory.
In U(1) we might expect renormalisation and screening, 
but not in the pure gauge theory.

Assuming the Ehrenfest identities to be true, we may infer 
$k^{\total}_\mu$ from a measurement of the derived field strength
operator, and break it down into some of its components. This allows
the consideration of lattices of a more physically interesting size.
In Table~\ref{tab_finsize} we see that the currents are remarkably
stable and free of finite volume effects even moving to lattices
large enough to support infrared physics. Finite volume effects on
the currents induced by a plaquette--sized source would thus appear
to be slight.

The Gribov ambiguity was neglected throughout. The
Morse functional $R[U]$ typically has a number of local maxima, each
giving a different set of maximally Abelian gauge fields corresponding
to the Gribov copies. There is some ambiguity regarding which of these 
to use and
one is usually selected randomly during gauge
fixing. The variation of gauge variant observables, such as the string
tension after Abelian projection, between these copies, although
not zero, is small enough that we may continue to neglect the ambiguity
in studying the properties  of this gauge
\cite{hart97}.
It is possible, however, that observables derived directly from the
Faddeev--Popov operator may be unduly sensitive to this ambiguity. Only
on lattices comparable in size to the confining length scale
($\sim 1$ fm) do we see nonperturbative effects, and it is on these 
lattices that $R[U]$ begins to have multiple maxima and Gribov copies
appear
\cite{hart97}.
The Ehrenfest identities were derived by infinitessimally (in
principle) shifting a configuration away from a maximum of $R[U]$, and
applying a corrective gauge transformation to return to the maximally
Abelian gauge. If the Gribov copies are not too numerous, it seems 
likely that the configuration does not forsake the attractive influence
of the original maximum, and thus returns to the `same' Gribov copy.
When the copies become very numerous, however, it may be that two
maxima of $R$ are sufficiently proximate that there is a `flat 
direction' between them, characterised by an extra (near) zero eigenvalue
in the Faddeev--Popov matrix.

A $4^4$ lattice at $\beta = 2.3$ is far too small to support
nonperturbative physics, and indeed the configurations we studied there 
exhibited only one maximum of $R$ in 1000 gauge fixings (we 
differentiated the maxima using the value of $R$ and the U(1) plaquette
action after Abelian projection, as in 
\cite{hart97}).
In none of our simulations were there any problems inverting the
Faddeev--Popov matrix, and we believe the effects of the Gribov ambiguity
to be very small. Only below $\beta \approx 2.15$ do Gribov copies
appear, and although we again had no apparent problems inverting, the
statistical noise prevents us from making any statements about the 
currents. On larger lattices, we have shown in Table~\ref{tab_finsize}
that the finite volume effects are slight, even on moving to physically
large lattices. This does not prove that the Gribov copies have no
effect (since we only undertook one gauge fixing per configuration), but 
the continuing smallness of the statistical errors
is perhaps indicative that that.

In a perturbative treatment of the gauge fixed action, the 
non--local Faddeev--Popov operator would be replaced by a Gaussian
integral over propagating ghost fields. In this sense, then, the
Faddeev--Popov current that we have isolated may be regarded as the
contribution of the ghost fields, hitherto little studied in the
maximally Abelian gauge. We conclude that in this context at
least these fields play little r\^ole since the Faddeev--Popov
current is only a fraction of the other currents.

\section{Summary}
\label{sec_summ}

In this paper we have exploited symmetries of the lattice partition
function to derive a set of exact, non--Abelian identities which
define the Abelian field strength operator and a conserved
electric current arising from the coset
fields traditionally discarded in Abelian projection. The current has
contributions from the action, the gauge fixing condition and the
Faddeev--Popov operator. Numerical studies on small
lattices verified the identity to within errors of a few per cent. 
We have found the Faddeev--Popov current in particular to be unusually 
sensitive to
systematic effects such as low numerical precision and poor random
number generators, but the origin of any remaining, subtle biases,
if they exist, is not clear; we have
already considered all terms in the partition function.

In a pure U(1) theory the static quark potential may be measured
using Wilson loops that correspond to unit charges moving in
closed loops, as demonstrated by 
$|\langle \diffn{\nu} f_{\nu \mu} \rangle| = \delta_W$.
In Abelian projected SU(2) the same measurements in the maximally
Abelian gauge yield an asymptotic area law decay and a string
tension that is only slightly less than the full non--Abelian
value. In other gauges it is not clear that an area law exists ---
certainly it is more troublesome to identify.

We have seen that in the context of the full theory the Abelian
Wilson loop must be reinterpreted. The coset fields renormalise 
the charge of the loop as measured by 
$|\langle \diffn{\nu} f_{\nu \mu} \rangle|$
and charge is also induced in the surrounding vacuum. Full
SU(2) has antiscreening/asymptotic freedom of colour
charge, and in the maximally Abelian gauge alone have we seen
analogous behaviour, in that the source charge is increased 
and induces charge of like polarity in the neighbouring vacuum.
Whether this renormalisation of charge can account for the
reduction of the string tension upon Abelian projection in 
this gauge is not clear. In other gauges, where Abelian
dominance of the string tension is not seen, the coset fields
appear to have a qualitatively different behaviour, acting
to suppress and screen the source charge.

In conclusion,
the improved field strength expression defined by the Ehrenfest 
identity does not coincide with the lattice version of
\cite{bernstein97}
of 't Hooft's proposed field strength operator
\cite{thooft74}.
The Abelian and monopole dominance of the string tension invites a
dual superconductor hypothesis for confinement. If this is to be
demonstrated quantitatively such as by verification of a (dual) London
equation then a a careful understanding of the field strength operator
is required. The Ehrenfest identities may provide this
\cite{hartprog}.
\vspace{0.20in}
\noindent {\bf Acknowledgements}

We thank M.I. Polikarpov and F. Gubarev for pointing out that we
should expect a contribution from the Faddeev--Popov determinant.
This work was supported in part by United States Department of Energy
grant DE-FG05-91 ER 40617.

\newpage
\def\baselinestretch{1.0}

\begin{table}
\begin{center}
\begin{tabular}{l*4{r@{.}l}}
\hline \hline
{\sc on the source} &
\multicolumn{4}{c}{$\beta = 2.3$} &
\multicolumn{4}{c}{$\beta = 2.5$} \\
& 
\multicolumn{2}{c}{`no gauge'} &
\multicolumn{2}{c}{MA gauge} &
\multicolumn{2}{c}{`no gauge'} &
\multicolumn{2}{c}{MA gauge} \\
\hline
\# meas. &
\multicolumn{2}{l}{10000} &
\multicolumn{2}{l}{10000} &
\multicolumn{2}{l}{10000} &
\multicolumn{2}{l}{20000} \\
\hline
$\beta
\la 
I_{\kappa \lambda}  . 
\left.
\frac{\partial S_W}{\partial \varepsilon}
\e
\ra$ &
0 & 0765 (3) & 0 & 6162 (8) & 0 & 0815 (2) & 0 & 68275 (58) \\
$k^{\static} . \la R_{\kappa \lambda} \ra $ &
0 & 0762 (1) & 0 & 5607 (5) & 0 & 0818 (1) & 0 & 63069 (21) \\
$\la 
I_{\kappa \lambda}  . 
k^{\gauge}
\ra $ &
\none & 0 & 0469 (1) & \none  & 0 & 04463 (5) \\
$\la 
I_{\kappa \lambda}  . 
k^{\fp}
\ra $ --- shift &
\none & 0 & 0063 (1) & \none  & 0 & 00565 (3) \\
$\la 
I_{\kappa \lambda}  . 
k^{\fp}
\ra $ --- corr. g.t. &
\none & 0 & 0034 (11) & \none  & 0 & 00133 (51) \\
\hline
\lhs eqn.~(\ref{eqn_ehren_test}) &
0 & 0003 (4) & -0 & 0011 (17) & -0 & 0003 (3) & 0 & 00045 (71) \\
\hline \hline
\end{tabular}
\end{center}
\caption{Numerical tests of the Ehrenfest identities on $4^4$ 
lattices, with and without gauge fixing.}
\label{tab_oldsource}
\end{table}

\begin{table}
\begin{center}
\begin{tabular}{l*4{r@{.}l}}
\hline \hline
{\sc on the source} &
\multicolumn{4}{c}{$\beta = 2.3$} &
\multicolumn{4}{c}{$\beta = 2.5$} \\
& 
\multicolumn{2}{c}{`no gauge'} &
\multicolumn{2}{c}{MA gauge} &
\multicolumn{2}{c}{`no gauge'} &
\multicolumn{2}{c}{MA gauge} \\
\hline
$k^{\dyn}$ &
-0 & 1266 (40) & 0 & 0800 (5) & -0 & 1293 (25) & 0 & 0781 (7) \\
$k^{\gauge}$ &
\none & 0 & 0836 (2) & \none  & 0 & 0711 (1) \\
$k^{\fp} $ &
\none & 0 & 0173 (20) & \none  & 0 & 0112 (3) \\
$k^{\static} + k^{\total}$ &
0 & 8734 (40) & 1 & 1809 (20) & 0 & 8707 (25) & 1 & 1604 (10) \\
\hline \hline
\\
\\
\hline \hline
{\sc off the source} &
\multicolumn{4}{c}{$\beta = 2.3$} &
\multicolumn{4}{c}{$\beta = 2.5$} \\
& 
\multicolumn{2}{c}{`no gauge'} &
\multicolumn{2}{c}{MA gauge} &
\multicolumn{2}{c}{`no gauge'} &
\multicolumn{2}{c}{MA gauge} \\
\hline
$k^{\dyn}$ &
-0 & 0074 (40) & 0 & 0139 (5) & -0 & 0126 (40) & 0 & 0145 (5) \\
$k^{\gauge}$ &
\none & 0 & 0042 (1) & \none  & 0 & 0037 (1) \\
$k^{\fp} $ &
\none & 0 & 0001 (25) & \none  & 0 & 0012 (3) \\
$k^{\static} + k^{\total}$ &
-0 & 0074 (40) & 0 & 0182 (25) & -0 & 0126 (40) & 0 & 0194 (10) \\
\hline \hline
\end{tabular}
\end{center}
\caption{Ehrenfest currents $\langle I_{\kappa \lambda}.k \rangle
/ \langle R_{\kappa \lambda} \rangle$ on $4^4$ 
lattices, with and without gauge fixing.}
\label{tab_onandoff}
\end{table}

\begin{table}
\begin{center}
\begin{tabular}{l*{3}{r@{.}l}}
\hline \hline
{\sc on the source} &
\multicolumn{2}{l}{$j^{\total}$} &
\multicolumn{2}{l}{$j^{\dyn}$} &
\multicolumn{2}{l}{$j^{\gauge} + j^{\fp}$}
\\
\hline
$\beta = 2.3$, $L=4$  & 0 & 1809 (20) & 0 & 0800 (5)  & 0 & 1009 (25) \\
$\beta = 2.3$, $L=6$  & 0 & 1776 (10) & 0 & 0674 (5)  & 0 & 1102 (5)  \\
$\beta = 2.3$, $L=8$  & 0 & 1735 (39) & 0 & 0642 (25) & 0 & 1093 (41) \\
$\beta = 2.3$, $L=10$ & 0 & 1840 (83) & 0 & 0654 (30) & 0 & 1186 (94) \\
\\
$\beta = 2.5$, $L=4$  & 0 & 1596 (7)  & 0 & 0781 (7)  & 0 & 0824 (13) \\
$\beta = 2.5$, $L=6$  & 0 & 1610 (15) & 0 & 0757 (17) & 0 & 0853 (17) \\
$\beta = 2.5$, $L=10$ & 0 & 1658 (40) & 0 & 0753 (16) & 0 & 0905 (35) \\
\hline \hline
\end{tabular}
\end{center}
\caption{Finite volume effects on the Ehrenfest
currents,
$\la I_{\kappa \lambda}  . k \ra_{gf} / 
\la R_{\kappa \lambda} \ra_{gf} $ 
in the maximally Abelian gauge.}
\label{tab_finsize}
\end{table}

\end{document}